# HistomicsML2.0: Fast interactive machine learning for whole slide imaging data

Sanghoon Lee[1], Mohamed Amgad[1], Deepak R. Chittajallu[2], Matt McCormick[2], Brian P Pollack[3,4], Habiba Elfandy[5], Hagar Hussein[6], David A Gutman[7], Lee AD Cooper[8]

[1] Weisberg Division of Computer Science, Marshall University, Huntington, WV USA
[2] Kitware, Inc., Clifton Park, NY USA
[3] Department of Dermatology, Emory University School of Medicine, Atlanta, GA USA
[4] Department of Pathology, Emory University School of Medicine, Atlanta, GA USA
[5] Department of Pathology, National Cancer Institute, Cairo, Egypt
[6] Department of Pathology, Cairo University, Cairo, Egypt
[5] Department of Neurology, Emory University School of Medicine, Atlanta, GA USA
[6] Department of Pathology, Northwestern University Feinberg School of Medicine, Chicago, IL USA

Correspondence: lee.cooper@northwestern.edu

## ABSTRACT

Extracting quantitative phenotypic information from whole-slide images presents significant challenges for investigators who are not experienced in developing image analysis algorithms. We present new software that enables rapid learn-by-example training of machine learning classifiers for detection of histologic patterns in whole-slide imaging datasets. HistomicsML2.0 uses convolutional networks to be readily adaptable to a variety of applications, provides a web-based user interface, and is available as a software container to simplify deployment.

## MAIN

Whole-slide digital imaging of histologic sections is increasingly commonplace in research. Multiresolution whole slide images (WSIs) are a potentially rich source of data for clinical and basic science investigations, however, extracting quantitative information from WSIs presents many challenges. Image analysis algorithms for detecting and classifying structures and patterns in WSIs are typically specific to a given tissue or dataset, and are difficult for investigators to develop or to adapt to new applications. The scale and format of whole slide images (WSIs) also prevents analysis with common image analysis software tools like ImageJ[1] or CellProfiler[2].

The emergence of learning based algorithms like convolutional neural networks (CNNs) has been transformational in biomedical image analysis. CNNs are largely powered by labeled data, and have been demonstrated capable of detecting, classifying, and delineating histological structures in digital pathology applications given adequate labeled datasets[3-4]. Given their reliance on training data and the increasing commodification of deep learning algorithms and software, CNNs can be readily adapted to new applications and utilized by a broader community who are capable of generating annotations but who may lack experience developing image analysis algorithms from scratch.

We developed HistomicsML2.0, a software platform for training and validation of CNN classifiers on WSI datasets. Training is performed using a learn-by-example approach where users interact with classification results and provide labels and corrections to prediction errors. During training an active learning algorithm helps users identify the most informative examples for labeling in order to minimize effort and increase prediction accuracy. HistomicsML2.0 uses superpixel segmentation to subdivide tissue regions into small patches that are then analyzed using a pre-trained CNN to extract features for classification. This combination of operations results in a fast and responsive algorithm that is scalable to WSI datasets and that can be adapted to classifying a wide variety of histologic structures and patterns.

Users are assisted in classifier training with an active learning approach that identifies informative examples in order to minimize user effort and increase prediction accuracy. The HistomicsML2.0 server software and data generation tools are provided as portable Docker software containers that serve user interfaces for training, validation, inference, and organizing datasets and that provide algorithms for data extraction from WSIs. In this paper we demonstrate how HistomicsML2.0 can be used to create accurate classifiers of tumor infiltrating lymphocytes in triple-negative carcinoma of the breast and cutaneous melanoma, and show that active learning training significantly improves accuracy.

This work builds on our prior system, HistomicsML, where users provide their own image segmentation data and feature extraction algorithms[5]. Generating this data requires technical expertise in image analysis and new algorithms must be developed and tuned for each application. This step has been eliminated in HistomicsML2.0 by using a more generalizable CNN patch classification approach that can be applied to a wide variety of applications and by users who are not primary developers of image analysis algorithms. Microscopy image analysis tools including ImageJ and CellProfiler provide user interfaces for interactive analysis of small images, but these tools are not equipped to visualize or analyze WSIs. CDeep3M[6] provides cloud-based CNN analysis capabilities for large microscopy datasets, but does not support WSIs and provides only a command-line interface.

An overview of the HistomicsML2.0 analysis pipeline is presented in Figure 1. Datasets are prepared for analysis by first performing a superpixel segmentation and feature extraction using the data generation Docker (Figure 1A). Superpixel segmentation is first performed at low magnification to generate a grid of small patches covering tissue regions. These superpixels adapt their boundaries to coincide with strong gradients like the edges of cell nuclei or tissue interfaces, producing an adaptive grid where the content of each patch is more homogeneous. A feature map describing each superpixel is generated in 3 steps: (1) A square patch centered on the superpixel is extracted at high magnification (2) A pre-trained VGG-16 CNN transforms this patch into a high-dimensional feature map (3) The size of this feature map is reduced using principal component analysis. The final step helps to improve speed during interactive classification by minimizing computation and storage. Extended details of our design and parameters used for experiments are presented in Methods and **Figure S1**.

During training, users navigate the dataset using prediction heatmaps superimposed over the WSIs, and provide corrective feedback to iteratively improve the classifier (**Figure 1B**). These heatmaps represent the classification uncertainty of the individual superpixels and help users to focus their labeling in regions that are enriched with superpixels that confound the classifier, and to avoid labeling of redundant superpixels that are familiar to the classifier. Labeling is performed by clicking individual superpixels or by using a cursor paint tool for labeling larger areas. Each labeled superpixel is used to update a neural network (NN) that classifies the superpixels. A patch is first extracted from the labeled superpixel and rotated duplicates of this patch are generated to increase the available training data. Low dimensional feature vectors are extracted from these augmented patches and used to update the NN. Finally, the updated NN is applied to the entire dataset to generate new predicted superpixel labels and uncertainty heatmaps.

This system is provided as a portable software container that can readily be deployed on server infrastructure using Docker Engine. This container serves user interfaces for training classifiers, reviewing training datasets, creating validation datasets, collaborative review, managing datasets and performing inference (see **Figure S2**). Trained classifiers are saved on the server and can be applied to new datasets to generate classification mask images for WSIs or a spreadsheet of statistics on the abundance of each class in each slide. Mask images can be used as inputs to other image analysis software, and the statistics can be used to perform correlative analyses with clinical outcomes or with genomic data.

We applied HistomicsML2.0 to develop classifiers of lymphocytic infiltration (TIL) in (1) triple-negative breast carcinomas (BRCA) and (2) primary cutaneous melanoma (SKCM). These experiments used WSIs of hematoxylin and eosin stained formalin-fixed paraffin embedded sections from The Cancer Genome Atlas (see **Table S1**). Classifier predictions were validated using ground truth annotations of tissue regions generated using the Digital Slide Archive[7] and a public dataset of breast cancer histopathology annotations dataset[8]. Accuracy was measured at the individual pixel resolution on annotated regions of holdout WSIs not used in training. Pixel resolution accuracy accounts for discretization errors due to superpixel boundaries not conforming to ground truth. All clinical data used in correlative analyses was taken from the TCGA Pan-Cancer resource[9].

Classification results are presented in **Figures 2A** and **2B**. The accuracy of the BRCA-TIL classifier trained with HistomicsML2.0 is 89.9% with 1700 superpixels labeled in 49 iterations. With only 500 labeled superpixels, this classifier is already 89.8% accurate in classifying TILs (**Figure 2C**, **Table S2**). To measure the benefit of active learning we generated ten BRCA-TIL classifiers by randomly selecting labeled superpixels from ground truth annotations of the training slides. These classifiers trained without active learning had consistently lower prediction accuracy (median 82.5% +/- 0.7). We found agreement between percentage of tissue corresponding to TILs with the percentages from ground truth annotations in testing images (Pearson ρ=0.68, see **Figure 2D**, **Table S3**). We also assessed the prognostic value of TIL percentages predicted by our HistomicsML2.0 classifier on testing WSIs from our triple-negative BRCA cohort (See **Figure 2E**, **Table S4**). Stratification of cases into TIL-rich and TIL-depleted using HistomicsML2.0 predictions accurately predicts disease progression risk in this cohort (logrank p=7.56e-3). The SKCM-TIL classifier converged to 92.4% accuracy (see **Figure 2F**). Analyses of clinical outcomes and comparisons to random classifiers were not possible for the SKCM-TIL predictions due to lack of ground-truth annotations.

HistomicsML2.0 was created for users who lack technical background in image analysis. The use of software containers that encompass all dependencies helps make deployment easy, avoiding the need for compilation or management of software library version conflicts. The system has been fully documented (https://histomicsml-taf.readthedocs.io/en/latest/index.html) and is available as an open-source tool for those who want to extend or modify its capabilities (https://github.com/CancerDataScience/HistomicsML-TA). While this software was designed and optimized to minimize computational requirements, a GPU equipped server will accelerate data-generation and enhance the responsiveness of the system during interactive training (see **Figure S3**). These resources are widely available as cloud services if local hardware is not available.

Our experiments provide evidence that HistomicsML2.0 can provide accurate classification of patterns in WSI datasets, and can yield valuable measurements for basic science or clinical investigations. The utilization of superpixel-based segmentation and CNN feature extraction allows HistomicsML2.0 to be applied in a wide variety of problems. The interfaces and active learning technology help to produce accurate classifiers while minimizing labeling effort, allowing users to rapidly explore WSI datasets and generate quantitative measurements of histology.

## REFERENCES

1. Schneider, C.A., Rasband, W.S. and Eliceiri, K.W. NIH Image to ImageJ: 25 years of image analysis. *Nature methods* **9**, 671 (2012).
2. Jones, T.R., Kang, I.H., Wheeler, D.B., Lindquist, R.A., Papallo, A., Sabatini, D.M., Golland, P. and Carpenter, A.E. CellProfiler Analyst: data exploration and analysis software for complex image-based screens. *BMC bioinformatics* **9**, 482 (2008).
3. Litjens, G., Sánchez, C.I., Timofeeva, N., Hermsen, M., Nagtegaal, I., Kovacs, I., Hulsbergen-Van De Kaa, C., Bult, P., Van Ginneken, B. and Van Der Laak, J. Deep learning as a tool for increased accuracy and efficiency of histopathological diagnosis. *Scientific reports*, **6**, 26286 (2016).
4. Bejnordi, B.E., Veta, M., Van Diest, P.J., Van Ginneken, B., Karssemeijer, N., Litjens, G., Van Der Laak, J.A., Hermsen, M., Manson, Q.F., Balkenhol, M. and Geessink, O. Diagnostic assessment of deep learning algorithms for detection of lymph node metastases in women with breast cancer. *Jama*, **318**, 2199-2210 (2017).
5. Nalisnik, M., Amgad, M., Lee, S., Halani, S.H., Vega, J.E.V., Brat, D.J., Gutman, D.A. and Cooper, L.A. Interactive phenotyping of large-scale histology imaging data with HistomicsML. *Scientific reports*, **7**, 14588 (2017).
6. Haberl, M.G., Churas, C., Tindall, L., Boassa, D., Phan, S., Bushong, E.A., Madany, M., Akay, R., Deerinck, T.J., Peltier, S.T. and Ellisman, M.H. CDeep3M—Plug-and-Play cloud-based deep learning for image segmentation. *Nature methods*, **15**, 677 (2018).
7. Gutman, D.A., Khalilia, M., Lee, S., Nalisnik, M., Mullen, Z., Beezley, J., Chittajallu, D.R., Manthey, D. and Cooper, L.A., 2017. The digital slide archive: A software platform for management, integration, and analysis of histology for cancer research. *Cancer research*, **77**, 75-78 (2017).
8. Amgad, M., Elfandy, H., Khallaf, H.H., Atteya, L.A., Elsebaie, M.A., Elnasr, L.S.A., Sakr, R.A., Salem, H.S., Ismail, A.F., Saad, A.M. and Ahmed, J. Structured Crowdsourcing Enables Convolutional Segmentation of Histology Images. Bioinformatics. (2019).

9. Liu, J., Lichtenberg, T., Hoadley, K.A., Poisson, L.M., Lazar, A.J., Cherniack, A.D., Kovatich, A.J., Benz, C.C., Levine, D.A., Lee, A.V. and Omberg, L., 2018. An integrated TCGA pan-cancer clinical data resource to drive high-quality survival outcome analytics. *Cell*, **173**, 400-416 (2018).
10. Akbani, R., Akdemir, K.C., Aksoy, B.A., Albert, M., Ally, A., Amin, S.B., Arachchi, H., Arora, A., Auman, J.T., Ayala, B. and Baboud, J., 2015. Genomic classification of cutaneous melanoma. *Cell*, **161**, 1681-1696 (2015).

## METHODS

**Normalization and superpixel segmentation.** Tissue areas in WSIs were masked using the HistomicsTK *simple_mask* function (https://digitalslidearchive.github.io/HistomicsTK/histomicstk.segmentation.html) to define the regions for data extraction and analysis. Masked tissue areas were color normalized using the *reinhard* function from the preprocessing module of HistomicsTK (https://digitalslidearchive.github.io/HistomicsTK/histomicstk.preprocessing.html). Normalized color images were then segmented with the scikit-image SLIC superpixel algorithm at 40X objective creating roughly 64 x 64 pixel sized superpixels. Square patches sized 128 x 128 pixels are extracted from the superpixel centroids and resized to 224 x 224 to be used as the input of the pre-trained VGG-16 model. These steps were performed on 8192 x 8192 tiled versions of the WSI. The objective magnifications for superpixel analysis and patch extraction as well as the superpixel parameters are fully adjustable.

**Generating convolutional feature maps.** The VGG-16 pre-trained CNN was used to extract feature maps from the color normalized patches (https://keras.io). This network consists of 13 convolutional layers, 5 max pooling layers and 3 fully-connected layers. We truncated this network to extract 4096-dimensional feature maps after the first fully-connected layer. These vectors are reduced to 64-dimensional feature maps using PCA to improve the storage footprint of datasets and responsiveness of the training interface, and to reduce inference time when applying trained classifiers to new datasets.

**Image and clinical data.** WSIs used in the experiments were obtained from The Cancer Genome Atlas (TCGA). All WSIs used in these experiments were derived from formalin fixed paraffin embedded sections stained with hematoxlyin and eosin. WSIs were reviewed to eliminate those containing artifacts affecting a significant portion of the tissue area including tissue folds, bubbles, poor focus, and scanning artifacts. Clinical data for the BRCA slides was obtained from the TCGA Pan-Cancer Survival Resource[9]. Progression-free interval outcomes were used as recommended by the survival resource. Triple-negative status was determined from the clinical data files obtained from the Genomic Data Commons (https://gdc.cancer.gov/). SKCM slides were limited to primary melanoma, where skin was the biopsy site, and limited to stages I and II to avoid inclusion of regional spread (lymph nodes) in the diagnostic slides. This stratification was similar to that reported in supplementary table 3[10].

**Ground truth annotation data.** Annotation data for the BRCA experiments was obtained from a public repository of pathologist generated and reviewed annotations[8]. This dataset contains freehand annotations of lymphocytic infiltration, tumor, stromal, and necrotic tissue regions performed within rectangular regions in WSIs from 149 triple-negative carcinomas.

Ground truth annotations for the SKCM experiments were generated using the Digital Slide Archive[7] and similar annotation protocols as those published in the BRCA annotation dataset[8]. Large rectangular regions of interest (ROIs) ranging from 2.4 - 11.2 mm$^2$ were selected for annotation and within each ROI, areas containing lymphocytic infiltration of the tumor-associated stroma were manually delineated using a free-hand polygon tool. For the purpose of this work, we considered all small mononuclear cells with lymphocyte-like morphology to be TILs. Dense purely plasma cell infiltrates were not included in TIL regions, but admixtures of lymphocytes and plasma cells were counted as lymphocytic infiltrates. Artifacts, normal epidermal or dermal structures, and nearby lymphoid aggregates were excluded from the regions of interest. The manual polygonal boundaries were stored on a Mongo database, and queried and converted to masks for analysis (pixel values encode region membership). This procedure follows the protocol described in the BRCA annotation dataset[8].

**Validation.** Slides in the BRCA and SKCM datasets were assigned to non-overlapping training and testing sets so that no single case/sample has slides represented in both sets. The BRCA dataset was split into training (45 slides) and testing (10 slides) sets. The SKCM dataset was split into training (40 slides) and testing (10 slides) sets. HistomicsML2.0 classifiers were trained on the training slides and their performance was evaluated on the testing slides. Since HistomicsML2.0 trains classifiers iteratively, classifier performance was evaluated at every iteration to analyze sensitivity to training set size.

Prediction accuracy of trained classifiers was measured on a pixel-wise basis. Superpixel predictions were mapped by filling each superpixel with the predicted class or class-probability. These prediction images were compared to ground truth annotations, measuring error on a pixel-wise basis to account for the fact that superpixel boundaries do not conform perfectly to ground truth annotations. Accuracy and area under curve (AUC) were measured by pixel-wise comparison of HistomicsML2.0 predictions for superpixels with masks from the ground truth annotations. Accuracy is defined as:

$$\text{Accuracy} = \frac{TP + TN}{P + N},$$

where $P$ and $N$ are the number of predicted positive and negative pixels respectively, $TP$ is the number of true positive pixel predictions and $TN$ is the number of true negative pixel predictions. AUC was computed using the receiver operating characteristic curve:

$$\text{TPR} = \frac{TP}{P}, \quad \text{FPR} = \frac{FP}{N},$$

where TPR is the true positive rate and FPR is false positive rate, and FP is the number of false positive pixel predictions.

**Training.** The HistomicsML2.0 BRCA classifier was trained over 49 iterations to label a total of 1741 superpixels. The HistomicsML2.0 SKCM classifier was trained over 56 iterations to label 2009 superpixels. A superpixel was considered TIL-positive if it contains one or more complete TIL nuclei. The model was initially primed by choosing 4 positive and 4 negative superpixels. After priming, each training cycle consisted of the following steps: 1. Choose a slide for analysis; 2. Use the uncertainty heatmap to navigate to a field of view with low prediction confidence; 3. View predicted superpixel labels in a high-magnification field of view and correct a roughly-equal number of false-positive and false-negative superpixels; 4. Repeat step 3 for two more fields of view in variable geographic locations, until the total number of corrected superpixels ranges between 30-60; 5. Retrain the network; 6. Navigate to next slide; 7. Repeat the process, making sure that all training slides are eventually represented in the labeled training set. Accuracy on the testing slides was not evaluated during the training process.

For the BRCA dataset, a series of 10 classifiers was generated as a comparison to evaluate the benefit of active learning training in HistomicsML2.0. Each training slide in the BRCA dataset has an associated ground truth annotation[8], and so we formed 10 training sets by sampling superpixels from these images and using the ground-truth annotations for labeling. These training sets were formulated to identically match the number of superpixels and slides where these superpixels were labeled during each iteration of HistomicsML2.0 classifier training. Superpixels were randomly sampled from annotation regions, and were labeled as positive if 75% or more of their pixels were annotated as TIL.

**Breast cancer data analysis.** We applied the trained HistomicsML2.0 classifier to an additional 96 slides to analyze TIL abundance and to evaluate the prognostic accuracy of HistomicsML2.0 predictions. We combined these 96 slides with the 10 testing slides to form an analysis set. These analysis slides have associated ground truth annotations[8] and do not overlap with cases in the training set.

Percent area occupied by TIL regions was calculated on a pixel-wise basis within the ROI for each analysis set slide for both the ground-truth annotations and the HistomicsML2.0 predictions. The correlation between predicted and actual percentages was calculated using Pearson correlation.

To evaluate prognostic accuracy, we used predicted TIL percentages calculated over WSIs. The median TIL percentage on training slides (2.75%) was used as a threshold to define TIL-rich and TIL-depleted categories. Analysis slides were placed in these categories using predicted TIL percentages, and a Kaplan Meier analysis was performed to compare these two groups. The logrank test was also used for hypothesis testing to compare survival distributions of the TIL-rich and TIL-depleted analysis cases.

**Machine learning.** HistomicsML2.0 uses a multi-layer neural network for superpixel classification and interactive learning. Keras version 2.2.2 and Tensorflow version 1.12 were used to implement the machine learning algorithms in HistomicsML2.0. In our experiments this neural network was configured with three layers (containing 64-32-1 neurons). The hidden layer uses rectified linear activation function and dropout fraction of 30%, and the output layer uses a sigmoidal activation for class prediction. This network was optimized using cross entropy loss with the Adam optimizer. All of these parameters are adjustable by editing the networks.py source file.

**Hardware.** Experiments were performed using a two-socket server with 2 x 16 Intel Xeon cores, 256 GB memory, 54 TB network mounted storage, and two NVIDIA Telsa P100 GPUs.

**Software.** All source code and documentation is available on Github (https://github.com/CancerDataScience/HistomicsML-TA) and ReadtheDocs (https://histomicsml-taf.readthedocs.io/en/latest/index.html). Containerized versions of software for data generation and the HistomicsML2.0 server have been published on Docker hub (https://cloud.docker.com/u/cancerdatascience/repository/docker/cancerdatascience/histomicsml).

**Data availability.** Whole slide images were downloaded from The Cancer Genome Atlas at the Genomic Data Commons (https://gdc.cancer.gov).

# FIGURES

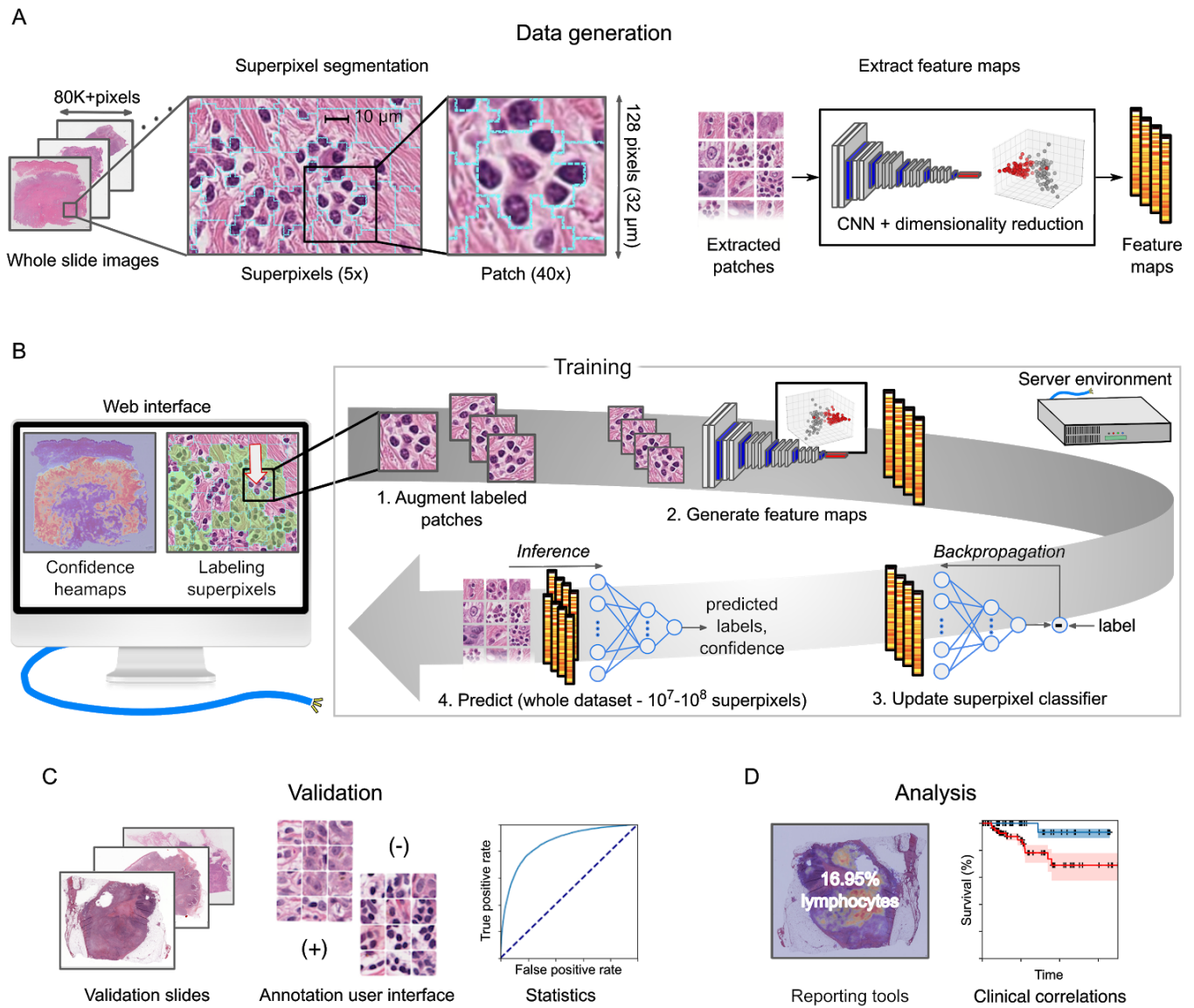

**Figure 1. Overview of HistomicsML2.0 software for whole-slide image analysis. (A)** Each WSI is analyzed at low magnification to divide the tissue into superpixel regions that adapt to gradients like cell nuclei borders or tissue interfaces. A patch is extracted from each superpixel at high-magnification and analyzed with a CNN to generate a feature map for machine learning. **(B)** A web interface communicating with a server is used to train a superpixel classifier. Zoomable WSI heatmaps guide users to regions that are enriched with superpixels having low prediction confidence. Each superpixel labeled by the user is augmented by rotation to increase the training set size, and feature maps from the labeled superpixels are used to update the classifier. The updated classifier is applied to the entire dataset to generate new predictions and confidence heatmaps, and the training process is repeated until the user is satisfied with the classifier predictions. **(C)** A validation interface helps users develop validation sets, to share and review these sets with others, and to calculate accuracy statistics for trained classifiers. **(D)** HistomicsML2.0 provides reporting tools to output summary statistics on WSI datasets. This data can be used to explore clinical or molecular correlations of histology.

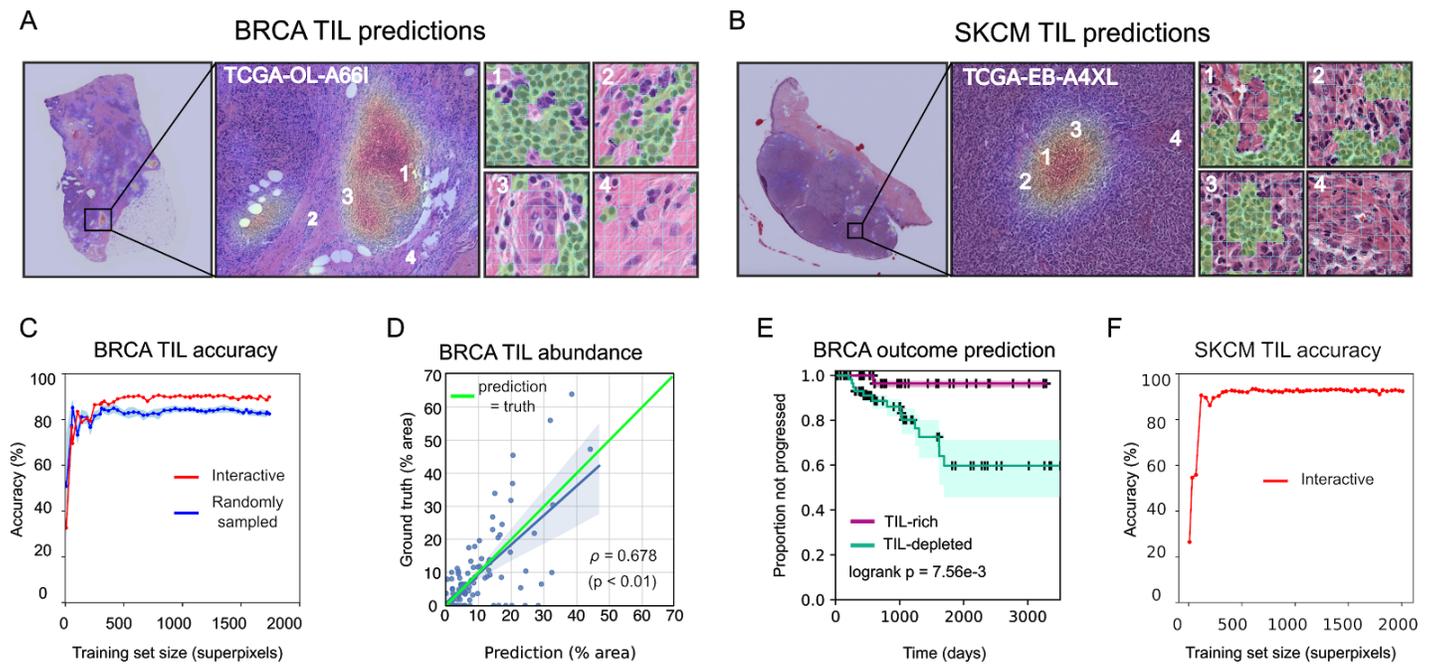

**Figure 2. Prediction results on BRCA and SKCM. (A-B)** TIL-density heatmaps and predicted TIL superpixels (green). Red heatmap areas indicate higher predicted TIL density. **(C)** Accuracy of BRCA-TIL classifier trained using active learning in HistomicsML2.0 (red). Accuracy of classifiers trained from randomly samples superpixels shown for comparison (blue is average +/- 1 SD). **(D)** Correlation between predicted and ground truth abundance of TIL positive areas in triple-negative test slides. **(E)** Kaplan Meier plot of disease progression for TIL-rich and TIL-depleted triple-negative cases as stratified by HistomicsML2.0 predictions on WSIs. **(F)** SKCM-TIL classifier accuracy.

**SUPPLEMENTARY FIGURES**

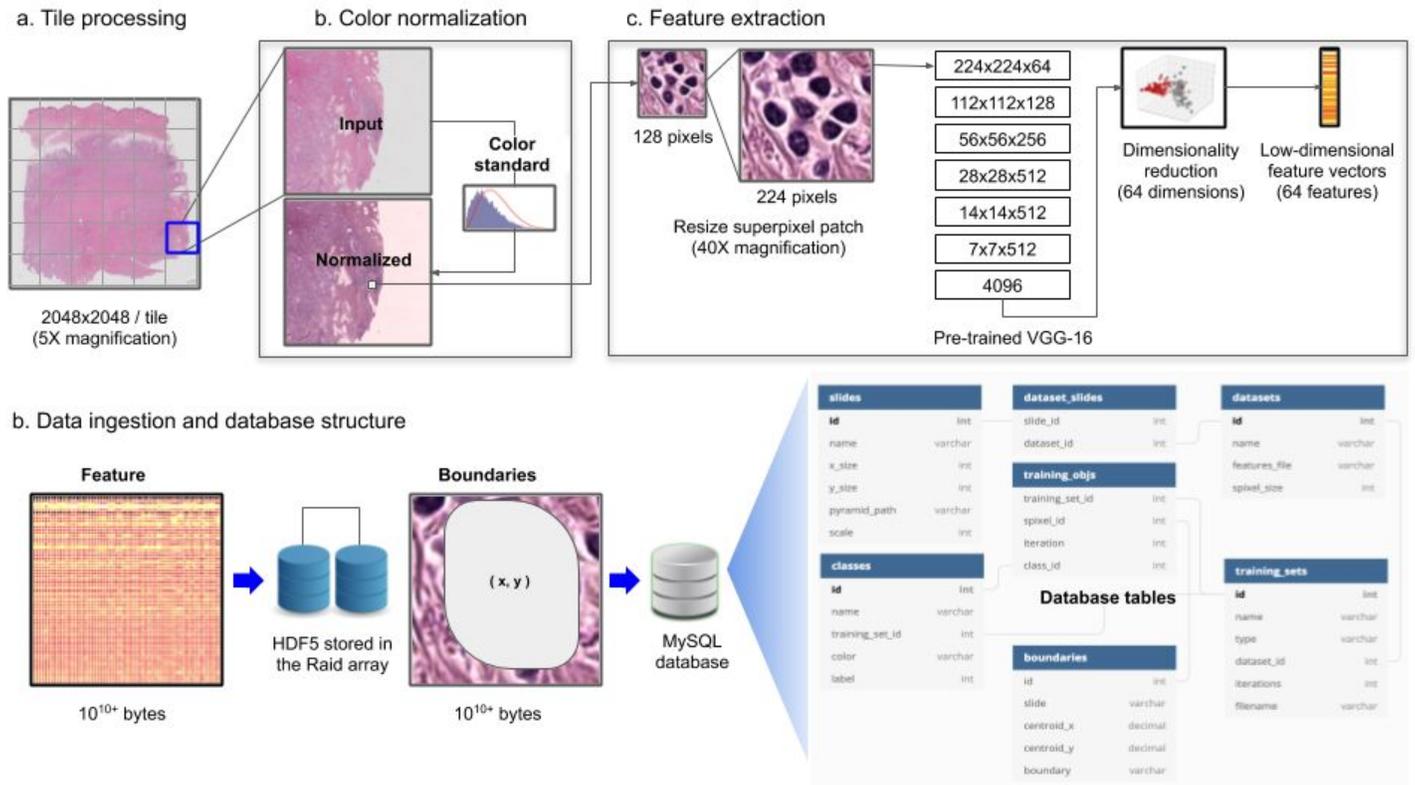

**Figure S1. Technical details of HistomicsML2.0 implementation.** (A) A WSI is partitioned into tiles and each tile is normalized to a standard hematoxylin and eosin slide with desirable color characteristics. Superpixel segmentation is performed and 128 x 128 patches are extracted at high-magnification and resized to 224 x 224. These patches are fed to a truncated pre-trained VGG-16 model extracting 4096-dimensional feature maps from the first fully connected layer. These feature maps are transformed to low dimensional feature maps using PCA (64 features in our experiments). Note that the superpixel sizing and the dimensionality of low-dimensional features are tunable parameters. All features are stored in HDF5 format and all boundaries are ingested into a MySQL database as a text file format. Additional database tables store information about training sessions, datasets, and whole-slide images.

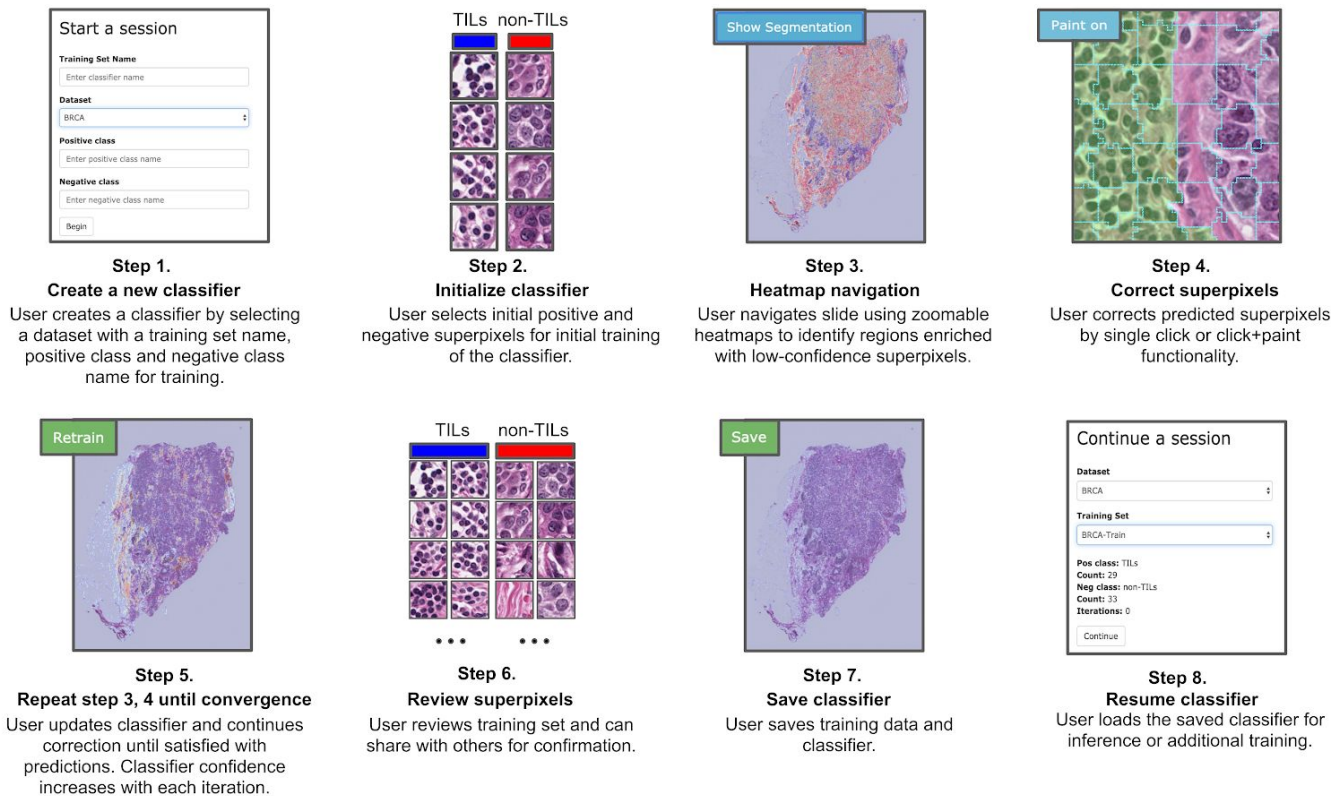

**Figure S2. Summary of training steps for HistomicsML2.0.** An instructional video is provided to help users navigate the menus.

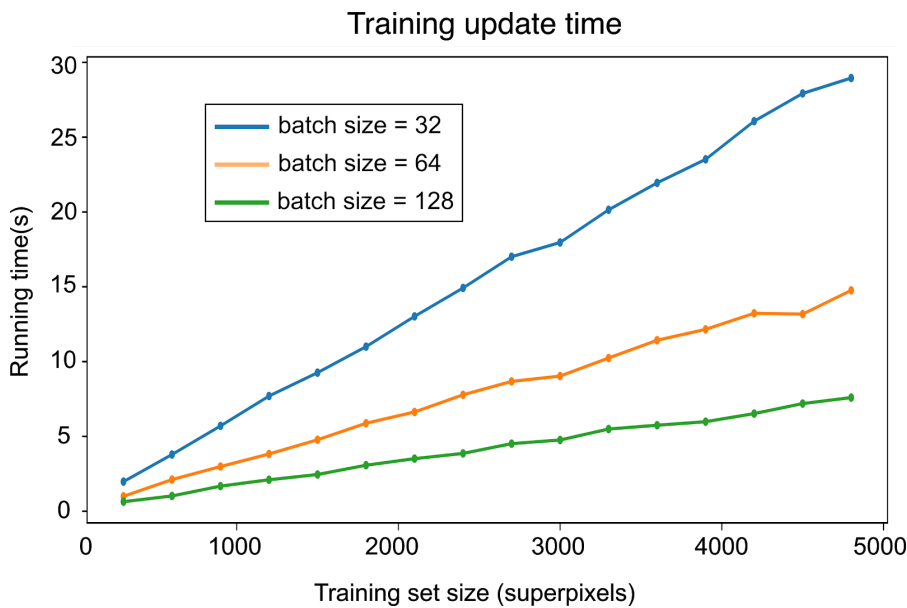

**Figure S3. Time to update neural network during training.** The time to update the network during training increases linearly as a function of the training set size. Larger batches decrease the number of communication sessions opened to the GPU and improve speed. Overall the update time remains below 30 seconds per update even for very large training sets. The average time required to perform inference on a dataset with 46 slides and 45 million superpixels is 23.47 seconds. Each iteration, requiring both a network update and full dataset inference, takes 30-35 seconds for most datasets. All experiments were performed on a server equipped with two NVIDIA P100 GPUs.

# SUPPLEMENTARY TABLES

See https://bit.ly/38QTcww for supplementary tables.

**Table S1.** Describes the slides used in experiments and their set assignment for validation experiments.

**Table S2.** Classification accuracy of classifiers trained with HistomicsML2.0 active learning and randomly sampled data. Prediction accuracy and training set size are reported for each training iteration for both BRCA and SKCM (Figures 2C, 2F). *Random sampling classifiers not available for SKCM dataset.

**Table S3.** BRCA abundance data (ground truth and predicted) for annotated ROIs (see Figure 2D).

**Table S4.** BRCA clinical data, TIL categories, and abundance (ground truth and predicted) for WSIs (see Figure 2D).